\documentclass[]{spie}  
\usepackage[]{graphicx}

\title{The Microchannel X-ray Telescope for the Gamma-Ray Burst mission SVOM} 

\author{D. G\"otz\supit{a}, J. Osborne\supit{b}, B. Cordier\supit{a}, J. Paul\supit{a}, P. Evans\supit{b}, A. Beardmore\supit{b}, A. Martindale\supit{b}, R. Willingale\supit{b}, P. O'Brien\supit{b}, S. Basa\supit{c}, C. Rossin\supit{c}, O. Godet\supit{d}, N. Webb\supit{d}, J. Greiner\supit{e}, K. Nandra\supit{e}, N. Meidinger\supit{e}, E. Perinati\supit{f}, A. Satangelo\supit{f}, K. Mercier\supit{g}, F. Gonzalez\supit{g}
\skiplinehalf
\supit{a}CEA Saclay - Irfu/Service d'Astrophysique (AIM), Orme des Merisiers, F-91191 Gif-sur-Yvette Cedex, France; \\
\supit{b}Department of Physics and Astronomy, University of Leicester, Leicester, LE1 7RH, UK; \\
\supit{c}LAM - Laboratoire d'Astrophysique de Marseille, P\^ole de l'\'Etoile Site de Château-Gombert, rue Fr\'ed\'eric Joliot-Curie 38, F-13388 Marseille Cedex 13, France;\\
\supit{d}Institut de Recherche en Astrophysique and Planétologie (IRAP), Université de Toulouse, UPS, 9 Avenue du colonel Roche, F- 31028 Toulouse Cedex 4, France;\\
\supit{e}Max-Planck-Institut f\"ur Extraterrestrische Physik, 85748 Garching, Germany;\\
\supit{f}IAAT, Sand 1, D-72076, T\"ubingen, Germany.;\\
\supit{g}CNES, 18 avenue Edouard Belin, F-31400 Toulouse, France
}

\authorinfo{Further author information: send correspondence to D. G\"otz; E-mail: diego.gotz@cea.fr}

\begin{document} 
  \maketitle 

\begin{abstract}
We present the Microchannel X-ray Telescope, a new light and compact focussing telescope that will be flying on the Sino-French SVOM mission dedicated to Gamma-Ray Burst science. The MXT design is based on the coupling of square pore micro-channel plates with a low noise pnCCD.
MXT will provide an effective area of about 50 cm$^{2}$, and its point spread function is expected to be better than 3.7 arc min (FWHM) on axis. The estimated sensitivity is adequate to detect all the afterglows of the SVOM GRBs, and to localize them to better then 60  arc sec after five minutes of observation.  
\end{abstract}



\section{Introduction}
\label{sec:intro}  

We present the Microchannel X-ray Telescope (MXT), a new telescope that will be flying on the Sino-French space mission SVOM dedicated to Gamma-Ray Burst (GRB) science. SVOM will carry a multi wavelength payload, which includes two instruments provided by France (ECLAIRs and MXT) and two provided by China (GRMs and VT). ECLAIRs is a wide field of view (2 sr) coded mask telescope, triggering in the 4 -- 250 keV energy range, and providing the initial GRB alerts. ECLAIRs is complemented by the GRMs, a set of three wide field non-imaging gamma-ray (30 keV -- 5 MeV) NaI spectrometers. Two narrow field instruments, a visible telescope, VT, and the MXT, will point the GRB location, after an autonomous slew of the satellite,  in order to study in detail the GRB afterglows, and to pinpoint the GRB locations. More details on the SVOM mission are given in Refs. \citenum{svom,mxt_mercier}.

The MXT is being developed under the responsibility of CNES Toulouse, in tight collaboration with CEA-Irfu, IRAP Toulouse, LAM Marseille, MPE Garching, and the University of Leicester. Its current design is based on the coupling of an optic made of square pore micro-channel plates (MCP), originally developed for the ESA BepiColombo mission to Mercury, with a low noise pnCCD, developed at MPE/HLL in the context of the \textit{DUO} mission concept and the eRosita instrument.
MXT is a compact and light ($<$30 kg) focussing X-ray telescope with a 1 m focal length, which will provide an effective area of about 50 cm$^{2}$ on axis. Its point spread function (PSF) is expected to be better than 3.7 arc min (FWHM), and its sensitivity ($\sim$2$\times$10$^{-12}$ erg cm$^{-2}$ s$^{-1}$ for a 10 ks observation, in the 0.2--10 keV energy range) adequate to detect practically all the X-ray afterglows of the SVOM GRBs.

MXT will be able to localize 90 (50)\% of SVOM GRB afterglows to better than 60 (20) arc sec (90\% confidence radius in the instrument reference frame) after five minutes of observation after the time of satellite stabilisation (about three minutes after trigger). In this paper we present the MXT design and expected performances based on simulations of a database of GRB afterglows observed by the X-ray Telescope (XRT), flying on the NASA Swift satellite\cite{swift}. They indicate that the SVOM GRB X-ray afterglows can be easily observed and studied with MXT from the temporal and spectral point of view up to 100 ksec after the gamma-ray event.

\section{Gamma-Ray Bursts and SVOM}

SVOM (Space-based multi-band astronomical Variable Objects Monitor) is a Sino-French mission dedicated to the detection, localization and study of Gamma Ray Bursts (GRBs), and other high-energy transient and variable phenomena (X-ray bursts, Soft Gamma Repeaters, Active Galactic Nuclei (AGN), Novae, ...). 
GRBs\cite{gehrels09} are short flashes of light - typically most easily detected as bursts of gamma- ray photons - coming from random directions of the sky at unpredictable times. Their gamma-ray spectra are non-thermal and their energy output usually peaks in the observer's frame in the range 50-500 keV.  Gamma-Ray Burst durations range from 0.001 s to about 1000 s or more, with a roughly bimodal distribution. Their gamma-ray light curves range from smooth, fast-rise and quasi-exponential decay, through curves with several peaks, to highly variable curves with many peaks.  GRBs are widely believed to be produced by powerful stellar explosions or compact objects mergers, that emit an ultra-relativistic jet in the observer direction. 

The incredible brightness of GRBs, which makes them visible across the entire Universe, and their connection with many fundamental physics questions make them invaluable tools for many branches of astrophysics and fundamental physics.
The prompt radiation of GRBs is followed by long lasting, but rapidly declining emission from radio to X-rays, the so-called ‘’afterglow’’ emission. A delayed long lasting emission has also been recently discovered at energies above 100 MeV thanks to the NASA Fermi satellite\cite{zhang11}. The rapid observation of the afterglows permits a derivation of accurate GRB localizations, owing to the better localization accuracy of X-ray and optical telescopes compared to what is possible with the very wide-field GRB detectors, provided that the GRB afterglow is observed sufficiently early when it is still bright. This precise localization is necessary to identify the host galaxy and/or measure the GRB redshift. 

The SVOM mission science objectives are summarized below:

\begin{itemize}

\item Permit the detection of all known types of GRBs, with a special emphasis on high-z
GRBs and nearby sub-luminous GRBs
\item Provide fast, accurate GRB positions
\item Measure the broadband spectral shape of the prompt emission (from visible to
MeV)
\item Measure the temporal properties of the prompt emission
\item Quickly identify the afterglows of detected GRBs, including those which are highly
red shifted (z $>$ 6)
\item Quickly provide arc second positions of detected afterglows, with redshift indicators
for detected GRBs
\end{itemize}

While GRBs observations will occupy only a small part of the SVOM observing time
(probably around 15\% at the beginning of the mission) the broad wavelength coverage
and good sensitivity of the on-board instrumentation allows for Non-GRB Science (NGS)
on selected topics during the remaining time. The constraint applicable to the NGS is that
the observation shall not influence the strategy of the main GRB goals. Before the launch
of SVOM, the SVOM Science Committee will define science topics that can be done
with SVOM instruments (including the SVOM Ground Follow-up Telescopes, GFTs) when
no GRB is being observed.

Some of the objectives presented above are directly related to the presence of an X-ray
telescope on board SVOM and to its performances, and have been summarized in the
the form of scientific requirements. The ones that
drive the design and scientific performances of the X-ray telescope are recalled here:
\begin{itemize}

\item \textit{SV-MRR-R5}: For at least 50\% of the localized GRBs, to point the GRB
direction and to allow at least 5 minutes of observation in the soft X-ray 
(down to 10$^{-11}$ erg cm$^{-2}$ s$^{-1}$) and optical bands (down to M$_{V}$=23).
\item \textit{SV-MRR-R7}: When a candidate counterpart of the GRB is detected in the soft X-ray
range, to measure its celestial coordinates with accuracy (90\% c.l. error radius) better
than 2 arc min at the detection threshold in the J2000 reference frame.
\end{itemize}

\section{MXT description}

Here we report the results of the MXT phase A design study. Some design optimizations will be studied and implemented during phase B.

The MXT is designed to detect and image X-ray photons in the soft X-ray energy range with good source location, time resolution, and spectroscopic capabilities. Its main goal is to significantly improve the GRB localizations derived on board by ECLAIRs and to permit spectral and timing studies of the afterglow emission.
The MXT telescope is currently based on the optical design of the MIXS-T telescope on board the ESA mission to Mercury BepiColombo, to be launched in 2016, coupled with a low noise X-ray camera (for more details on the MXT camera, see Ref.  \citenum{mxt_meuris}, for a full technical MXT description, see Ref. \citenum{mxt_mercier}).





\subsection{MXT Optics}

For the phase A study it has been assumed that the optics of the MXT will be a copy of that of MIXS-T. A short description is given here; for a detailed description of the MIXS-T telescope, see Ref. \citenum{fraser10}.

\begin{figure}[ht!]
   \begin{center}
   \includegraphics[width=15cm]{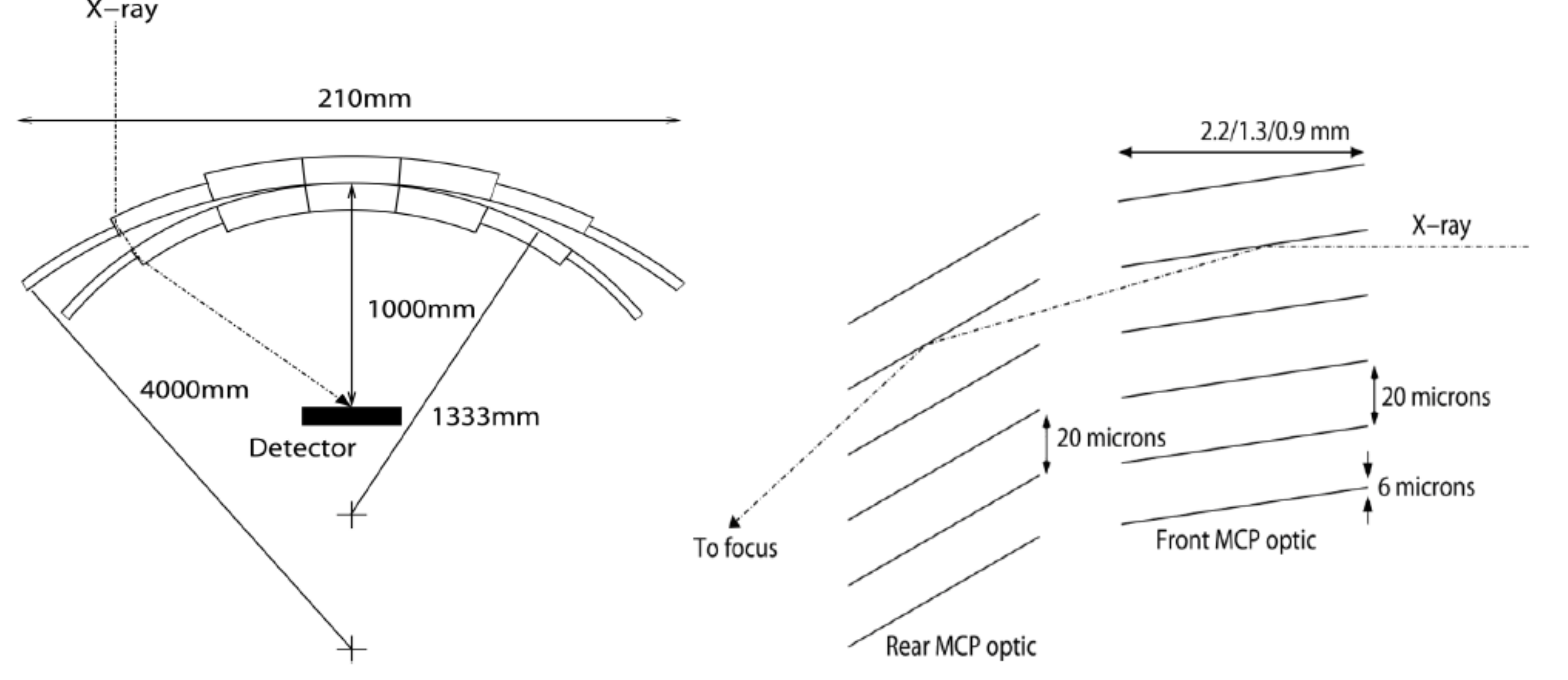}
   \end{center}
   \caption{Left: MIXS-T optical scheme. The two different slumping radii (4m and 1.33 m) give the focal length of 1 m. Right: optical path of an X-ray entering the MCPs and exiting after two reflections.}
	\label{fig:mcps1}
	\end{figure}

The optical design, which meets all the top-level scientific and sub-system requirements (see following sections), will make use of radially packed 20 micron square-pore micro channel plate optics, in a conical approximation to the Wolter type I focusing geometry, typically used in X-ray astronomy. The glass MCPs are coated with a thin layer of a high Z element (e.g. Platinum, Iridium, etc.) to boost reflectivity, especially at higher energies. The focal length is 1 m, determined by the 4 and 1.33 m slump radii of the front and the rear MCP plates. X-rays entering in the first plate are reflected at grazing incidence from an internal channel wall; upon exiting the front MCP, the X-ray enters a micro-channel in the rear MCP where a second reflection takes place. After two reflections, the beam converges in the focal plane (see Fig. \ref{fig:mcps1}). 
The 210 mm diameter optic is assembled from three rings of tandem pairs of MCPs (see Fig. \ref{fig:pnccd}), each of which is a sector of a circle, slumped to the figure of the surface of a sphere. Front and rear MCPs together form “tandems” which are arranged in three rings with different thickness (2.2 mm inner, 1.3 mm middle, and 0.9 mm outer) to approximate the ideal 1/r thickness profile which maximizes the throughput of the telescope by maximizing the probability of a single reflection in each MCP.
  
An electron diverter reducing the possibility of electrons reaching the detector is also part of the optics.


\subsection{MXT Detector}

\begin{figure}[ht!]
   \begin{center}
      \includegraphics[height=6cm]{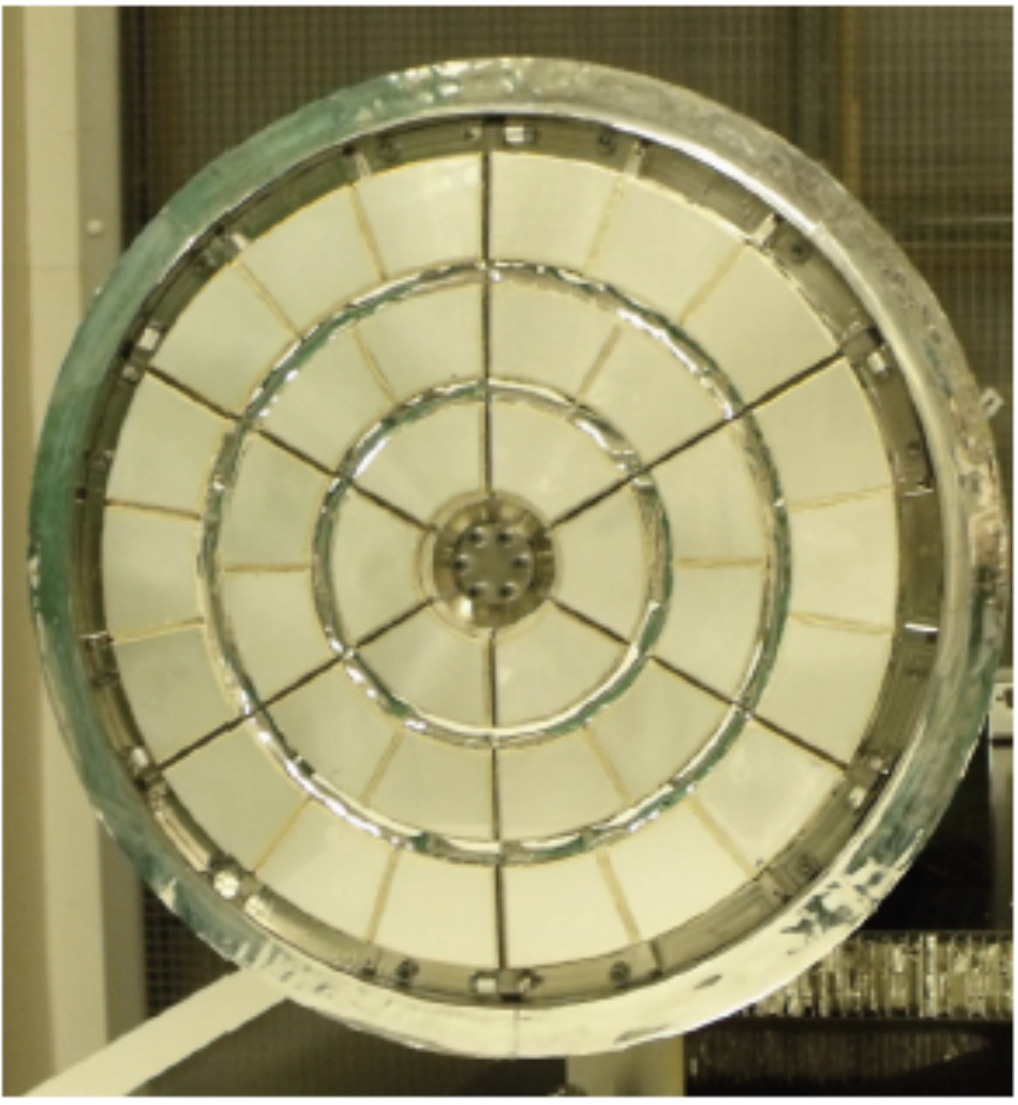}
      \hspace{1cm}
   \includegraphics[height=7cm]{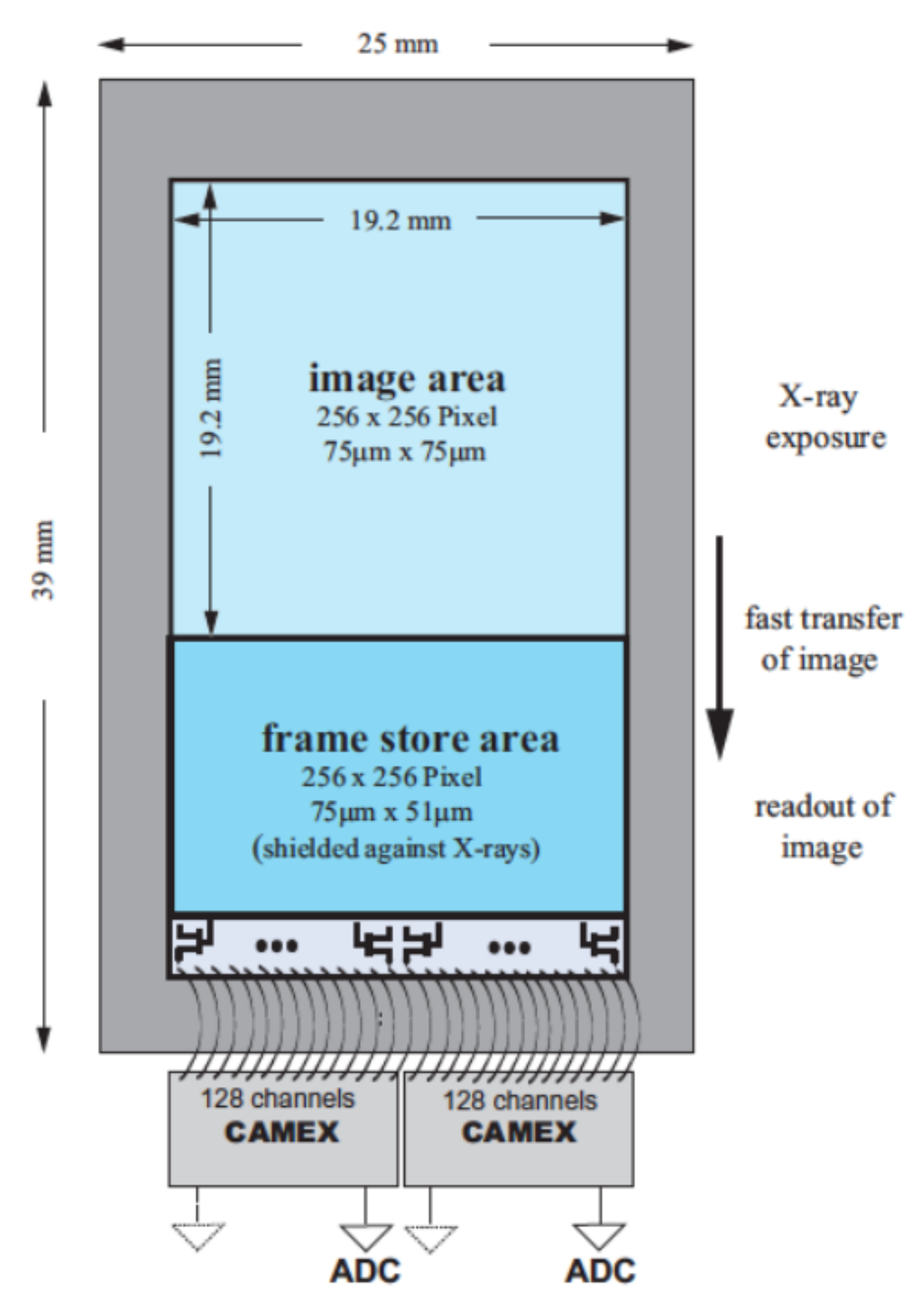}
   \end{center}
   \caption{Left: MIXS-T optics flight model. Right: Conceptual scheme of the \textit{DUO} pnCCD.}
	\label{fig:pnccd}
	\end{figure}

The detector at the focal plane of MXT will be provided by the Max Planck Institut f\"ur Extraterrestrische Physik (MPE), and will be a pnCCD, originally developed for the \textit{DUO} mission concept, and for the early phases of the eRosita instrument. A conceptual scheme of the pnCCD is shown in Fig. \ref{fig:pnccd}.

It is a frame transfer CCD with a 256x256 pixels imaging area. The pixel size is 75 $\mu$m, making a useful active area of 1.92$\times$1.92 cm$^{2}$, of fully depleted Si of 450 $\mu$m thickness. The frame store area has the same number of pixels, but with a reduced height of 51 $\mu$m, and each column is read out individually by two CAMEX chips, each one reading one half of the detector.
The transfer rate of the image to the frame store is very quick taking only 100 $\mu$s implying very few out-of-time (OOT) events. The minimum readout speed of the CCD is 6 ms. For more details on the \textit{DUO} pnCCD, see Ref. \citenum{meidinger06}.

\section{MXT Scientific Performance}
An accurate localization of the GRBs is essential to allow a successful search for an optical afterglow, for a redshift measure, and/or the identification of the host galaxy. After the GRB trigger and the satellite slew, the MXT must be able to cover with sufficient sensitivity the entire ECLAIRs error region, to detect the X-ray emission from the GRB afterglow, and to compute in near real time a first X-ray GRB afterglow position (which is immediately sent to the VT and to the ground via a VHF emitter). It must also be possible to derive one (or more) improved position(s) as more data are accumulated by the MXT, and, if the improved position is significantly better than the previous one(s), it will also be transmitted to the ground and to the platform. Ideally, MXT localizations should be obtained for the largest possible number of GRBs discovered by SVOM. Since the intensity of the X-ray afterglows decreases with time, this sets requirements at the satellite level (e.g. maximal duration of the slew to the ECLAIRs derived position, pointing stabilisation, etc.) as well as on the sensitivity of the MXT.

The MXT expected performance are summarized in Table \ref{tab:scireq}, and are described in some detail in the following sections.

\begin{table}[h]
\caption{MXT expected scientific performance.} 
\label{tab:scireq}
\begin{center}       
\begin{tabular}{|l|l|} 
\hline
Energy Range &	0.2-10 keV\\
\hline
Field of view &	64$\times$64 arc min\\
\hline
Point Spread Function &	3.7 arc min (FWHM @ 1.5 keV)\\
\hline
Sensitivity (5 $\sigma$) &	$\sim$10$^{-10}$ erg cm$^{-2}$ s$^{-1}$ in 10 s\\
 &  $\sim$2$\times$10$^{-12}$ erg cm$^{-2}$ s$^{-1}$ in 10 ks\\
\hline
Throughput &	1 mCrab  $\sim$0.20 ct/s  for N$_{H}$=4.5x10$^{21}$ cm$^{-2}$, photon index = 2.08\\
\hline
Energy Resolution &	$\sim$75 eV (FWHM @ 1 keV)\\
\hline
Time Resolution	& 100 ms\\
\hline 
\end{tabular}
\end{center}
\end{table}

\subsection{Energy Range}

\begin{figure}[ht!]
   \begin{center}
   \includegraphics[width=8cm]{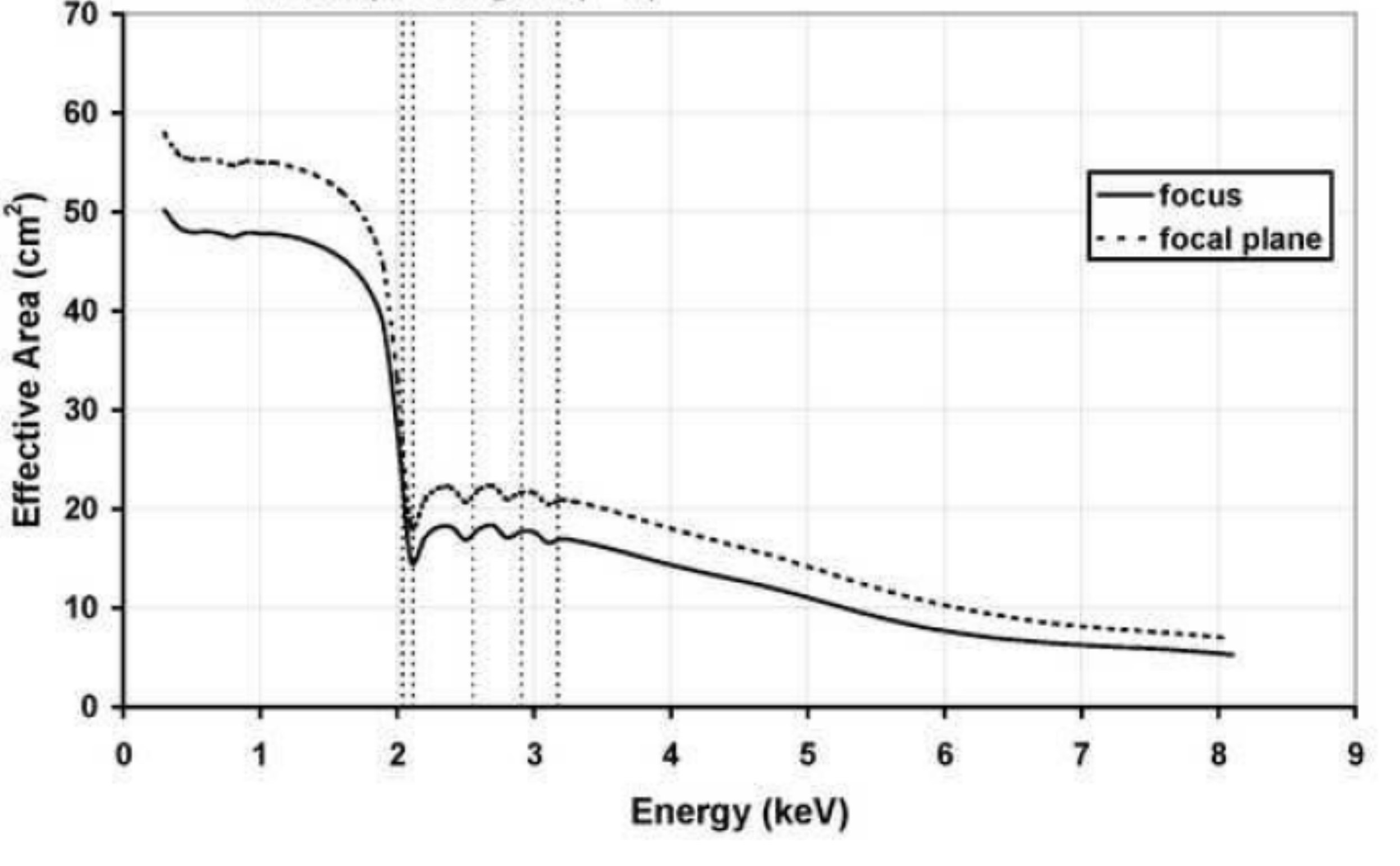}
   \includegraphics[width=7cm]{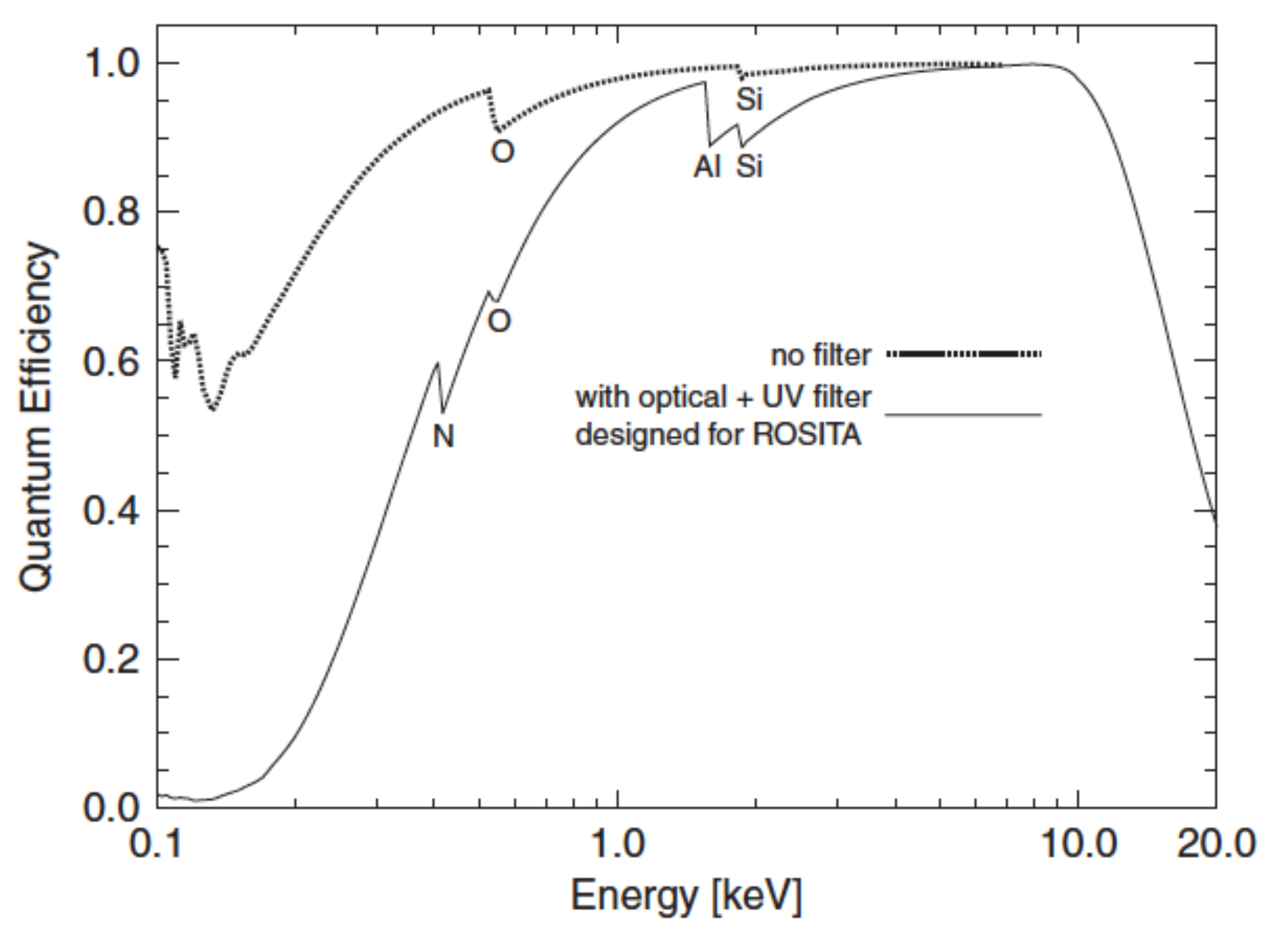}
   \end{center}
   \caption{Left: MIXS-T optics on axis effective area\cite{fraser10}. The continuous line refers to focussed X-rays, while the dashed one includes also the single reflection components (which are diffused over a large detector area). Right: Quantum efficiency of the frame store pnCCD. The upper curve represents the standard quantum efficiency of the pnCCD. The lower curve is the expected quantum efficiency with the optical/UV filter designed for the eROSITA instrument\cite{meidinger06}.}
	\label{fig:mcps2}
	\end{figure}

\begin{figure}[ht!]
   \begin{center}
   \includegraphics[width=8cm]{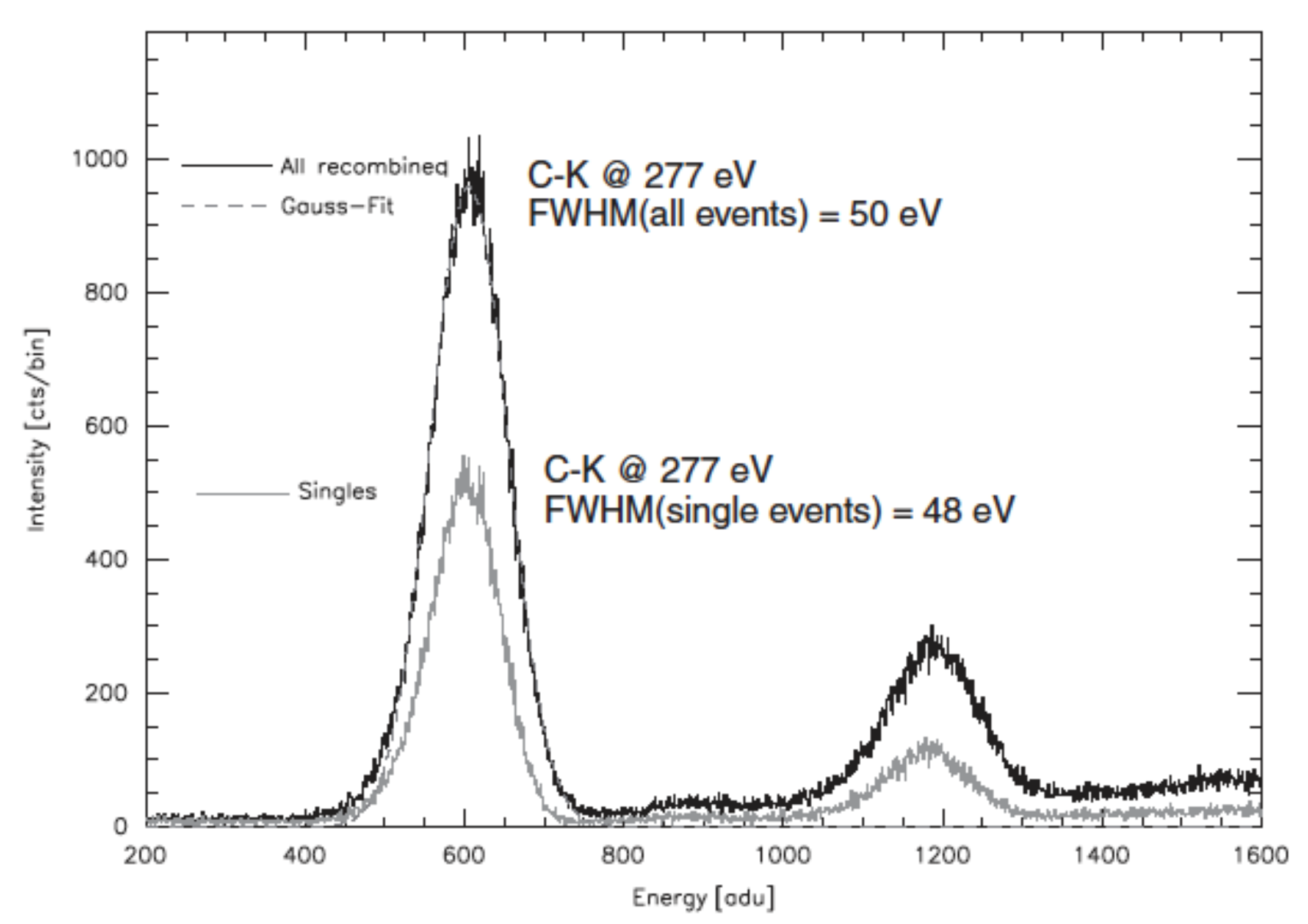}
   \includegraphics[width=8cm]{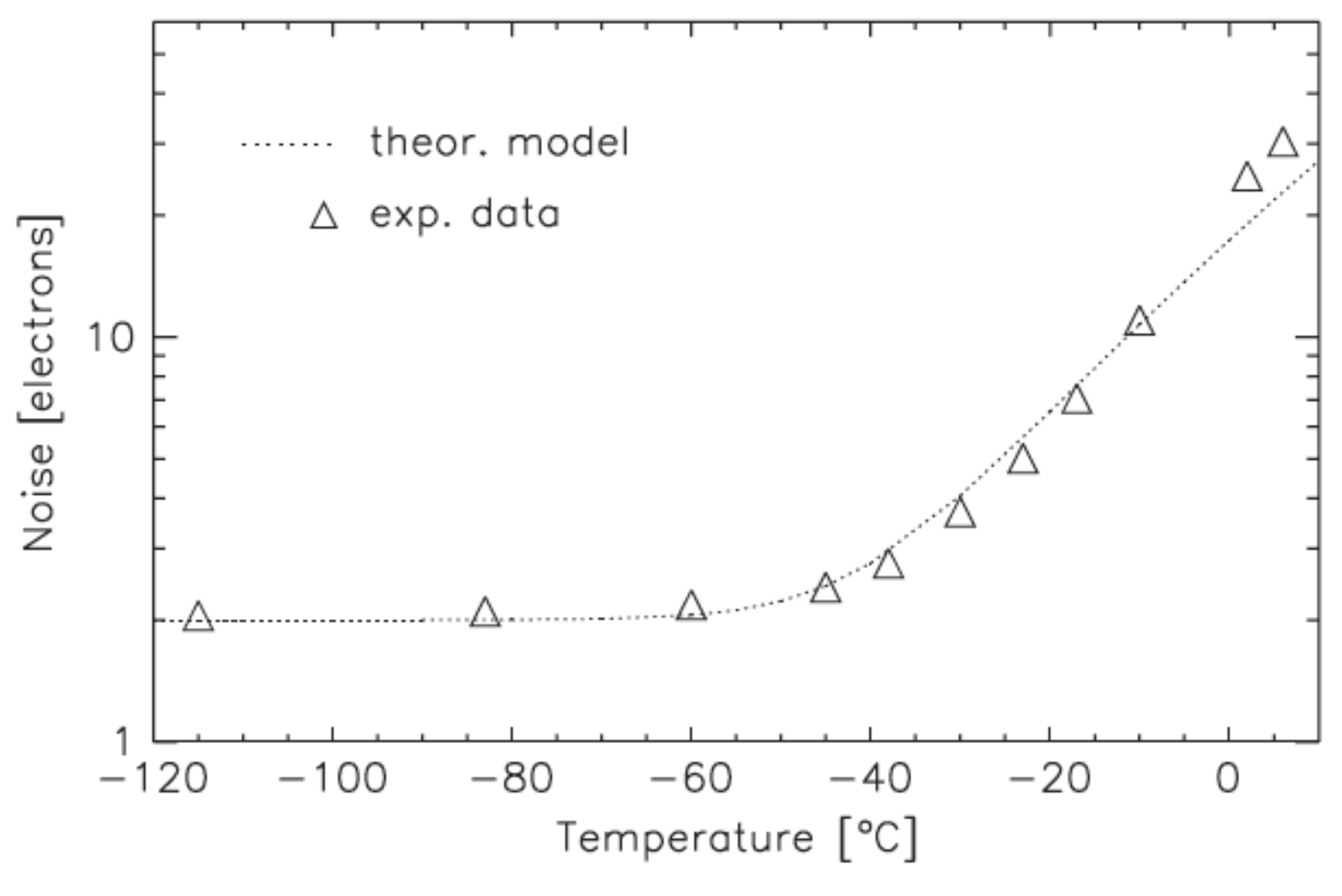}
   \end{center}
   \caption{Left: Low-energy response of the frame store pnCCD shown for a C–K spectrum measured at -40$^{\circ}$C with a frame rate of 125/s\cite{meidinger06}. The FWHM of the spectral line is ~50 eV. Right: Expected readout noise as a function of temperature\cite{meidinger06}.}
	\label{fig:pn}
	\end{figure}

The bulk of the X-ray photons during GRB afterglows is emitted below 1 keV, and are detected in this range even allowing for the partial absorption of the flux in our Galaxy and in the GRB host galaxy.
This fact combined with the expected effective area of the MIXS-T optics, which peaks below 1.5 keV, see Fig. \ref{fig:mcps2}, pushes the choice of a detector to have a low energy threshold, as low as possible. This can be obtained only by an accurate design of the detector and its associated electronics as well as the optical filters design, see later. This can be achieved in principle by the pnCCD which has a low readout noise of 2 e$^{-}$ rms at temperatures below $\sim$-65$^{\circ}$C, and has therefore very good spectral capabilities at low energies, see Fig. \ref{fig:pn}, and is adapted in providing a sufficiently low energy threshold, see Fig. \ref{fig:mcps2}.


On the other hand, despite the modest effective area above a few keV (see Section \ref{sec:sens}), some signal can be measured up to 10 keV with the pnCCD, thanks to its thickness (450 $\mu$m).

\subsection{Field of View and Angluar Resolution}
The geometrical field of view, defined by the MIXS-T optics and the physical size of the pnCCD is adequate in order to cover, including some margin, the whole GRB error circle obtained by the on-board localization with the ECLAIRs instrument, even for GRBs only slightly above the trigger threshold. In the current design, the largest error box (90\% c.l.) is expected to be 12 arc min (J2000) for ECLAIRs triggers (see Ref. \citenum{eclairs}). But this value could increase to 15 arc min, if an improved mask design of ECLAIRs is implemented, which at the expense of some loss in localization accuracy, could imply a gain of $\sim$30\% in the number of detected GRBs, relative to the current mask.  The pnCCD will provide a 64$\times$64 arc min FOV. This FOV is sufficiently large to cover the possible new ECLAIRs error box, and, if needed, also to point also “sub-threshold” GRBs. In addition the X-ray counterpart will enter earlier in a large FOV during the satellite slew, when the afterglow is brighter. This allows the study the physics of the end of the prompt emission, the very early afterglow, and provides the opportunity to localize faint afterglows (like typically the ones of short GRBs, and “naked” long GRBs) thanks to a drift correction algorithm (either on board or on ground).

The angular resolution of the MXT is driven by the combination of the MIXS-T PSF and its sampling by the detector pixels. The modelled value for the MIXS-T PSF used in this paper is based on early laboratory measurements of an incomplete MIXS-T optics. Its FWHM is 3.7 arc min
at 1.5 keV for off-axis angles smaller than 10 arc min. In fact the PSF width is not a strong function of the off-axis angle, and there is no significant degradation within 20 arc min off axis at least.
The PSF core is globally Gaussian, although there is a single reflection component to the PSF for off-axis sources which results in a faint asymmetric component.
The 75 $\mu$m pixels of the pnCCD correspond to 15 arc sec angular size, allowing the PSF to be fully sampled, even in the case of an improvement of the PSF down to 2 arc min FWHM, which is the goal for the MXT instrument. 
As shown in the following section, this PSF will yield sufficiently accurate positions (typically $<$ 2 arc min for 90\% of the bursts 10 minutes after detection) to allow the ground-based follow-up telescopes to observe efficiently the GRB afterglow. On the other hand, the goal performance will allow also larger facilities to start the follow-up early, in the case a sufficiently accurate pointing knowledge is available in real time to MXT.

\subsection{Sensitivity and Source Localization Accuracy}

The precision to which the MXT will be able to locate new sources (in its own reference frame) is dependant on the point spread function (PSF) of the optics, the size of the detector pixels, the number of counts used, the background count rate, and the ability of the algorithm used to make the measurement. This position has to be translated into the J2000 celestial reference frame, thus introducing additional dependencies on the transformations from the MXT to the spacecraft frame and spacecraft to star tracker J2000 frame (and in ground analysis, the spacecraft to the VT J2000 frame). 
The goal of providing an error box of the order of 2 arc min or smaller for the majority of the SVOM GRBs has been tested through a set of simulations. The input to these simulations is the Swift/XRT database of X-ray afterglows maintained at the University of Leicester by the XRT group. The XRT afterglow data have been modelled, and the models folded through the MXT response (derived by analogy from the XRT one), in order to calculate the expected count rate on the detector. 
The number of counts used by the MXT to determine a source position will depend on the source intensity (which is a strong function of time for GRBs), the source spectrum, the time interval of count accumulation, the efficiency with which the MXT records counts from incident photons (a strong function on energy), and the fraction of total counts used.
The source intensity as a function of time is taken from the Swift XRT afterglow catalogue, which contains 0.3-10 keV light curves of all Swift afterglows at 2.5 second time binning or better\cite{evans09}. The source spectrum is likewise from the XRT afterglow catalogue, each input XRT afterglow has its own XRT spectrum; for the purposes of simulation source spectra are assumed to have constant spectral slope and absorption throughout their evolution (slope changes are observed early in the afterglow evolution, but their net effect is minor). 
The SVOM GRB response time-scale is expected to be the same of that of Swift: 80\% of bursts are slewed to within 3 minutes, this simplifies the simulation of MXT count accumulation, as the time intervals to be used can be taken directly from the individual Swift XRT light curves (like Swift, SVOM will spend all of the early post-trigger orbits observing the afterglow, the pattern of visibility is assumed to be the same); this results in a median first orbit GRB observation duration of 1070 seconds. 
The different orbit inclination of Swift (21$^{\circ}$) and SVOM (30$^{\circ}$) results in a negligible difference in the useful time spent outside the South Atlantic Anomaly (Swift 88\%, SVOM 86\%).
  

The Swift XRT afterglows are from Swift BAT GRBs, while the GRBs detected by ECLAIRs are likely to be brighter on average due to the lower sensitivity of ECLAIRs, but softer on average due to its lower energy threshold than the BAT. The simulations made use of subsets of the Swift-detected GRBs representative of the ECLAIRs sample, that is the brightest 70\% of GRBs and the brightest 85\% of XRFs. In addition, the uncertainties in ECLAIRs positions will result in the bursts being placed off the MXT optical axis, assuming all bursts have 11 arc min 90\% c.l. errors (ECLAIRs SNR=8) results in a mean MXT off-axis angle of 6 arc minute and thus a vignetting factor of 0.79. The ability of MXT to convert incident photons to counts is limited at the low energy end by the transparency of the optical blocking filters, and at high energies mostly by the reflectivity of the 1 m focal length optic. The MXT effective area curve was created by use of the ray-traced optic reflectivity model (Ir coating assumed), the assumption of 200 nm of Al for filtering optical light, and the quantum efficiency of the e-Rosita back-illuminated CCD, which is similar to that of \textit{DUO}. With this, the median afterglow spectrum results in 0.43 MXT c/s per XRT c/s (a comparison between the XRT and MXT effective areas as a function of energy is shown in Fig. \ref{fig:sens}). Finally, a PSF fraction of 0.8 is assumed to be accumulated when determining a GRB position.


The last component of the calculation is the background count rate of the MXT. This is derived from the cosmic X-ray background\cite{moretti09}, the MXT internal background (i.e. the component due to the cosmic ray particles interaction with the detector, and to the electronic readout chain), and the non-focussed source X-rays. Among these, the CXB significantly dominates, giving 1$\times$10$^{-2}$ c/s/pixel (assuming the same GRB vignetting factor of 79\%), i.e. 0.021 c/s per 0.3-6 keV R$_{0.8}$=8.5 arc min beam; stray light X-rays are ignored in the simulations. 
The algorithm to be used by the MXT in determining positions on-board is assumed to be a perfect implementation of the best simple method available, the Iterative Weighted Centre Of Gravity method. This method has been shown to work as well as the ideal PSF-fitting approach, whilst using much fewer resources than a full PSF fit. An indication of PSF-likeness can be derived from the number of iteration needed or the measured width. 
90,000 Monte-Carlo simulations of IWCOG position determination with the PSF assumptions described above have shown that the 90\% confidence error radius in arc sec of an MXT position determination is:

R$_{90}$ = 411 C$^{-0.5}$ + 388.4 (C/B)$^{-1.929}$  [C/B $<$ 10.18]

R$_{90}$ = 411 C$^{-0.5}$ + 22.6 (C/B)$^{-0.7059}$  [C/B $>$ 10.18]

Where C is the source count and B is the background count. 

Making use of the assumptions described, and integrating counts up to given times for all GRBs results in cumulative distributions of 90\% confidence error radii as a function of time. 
The MXT is able to determine positions to within 6 arc sec or better in 10\% of cases within 15 minutes of an ECLAIRs trigger, and to within 51 arc sec  or better within 90\% of cases in 5 minutes; table \ref{tab:loc} shows some characteristic points in the position error distributions. These values are in the MXT reference frame and so are subject to additional error components in relating the position to the J2000 celestial frame. 

\begin{table}
\caption{Localization accuracy of all the simulated GRBs as a function of time.}
\label{tab:loc}
\begin{center}       
\begin{tabular}{|l|l|l|} 
\hline
Fraction of localised sources & Localisation Error (90\% c.l.)&	Time after ECLAIRs trigger\\
\hline
50\%	 & $<$ 17 arc sec	& 5 min\\
90\%	 & $<$ 51 arc sec	& 5 min\\
50\%	 & $<$ 15 arc sec	& 10 min\\
50\%	 & $<$ 14 arc sec &	15 min\\

\hline 
\end{tabular}
\end{center}
\end{table}

\begin{figure}[ht!]
   \begin{center}
   \includegraphics[width=15cm]{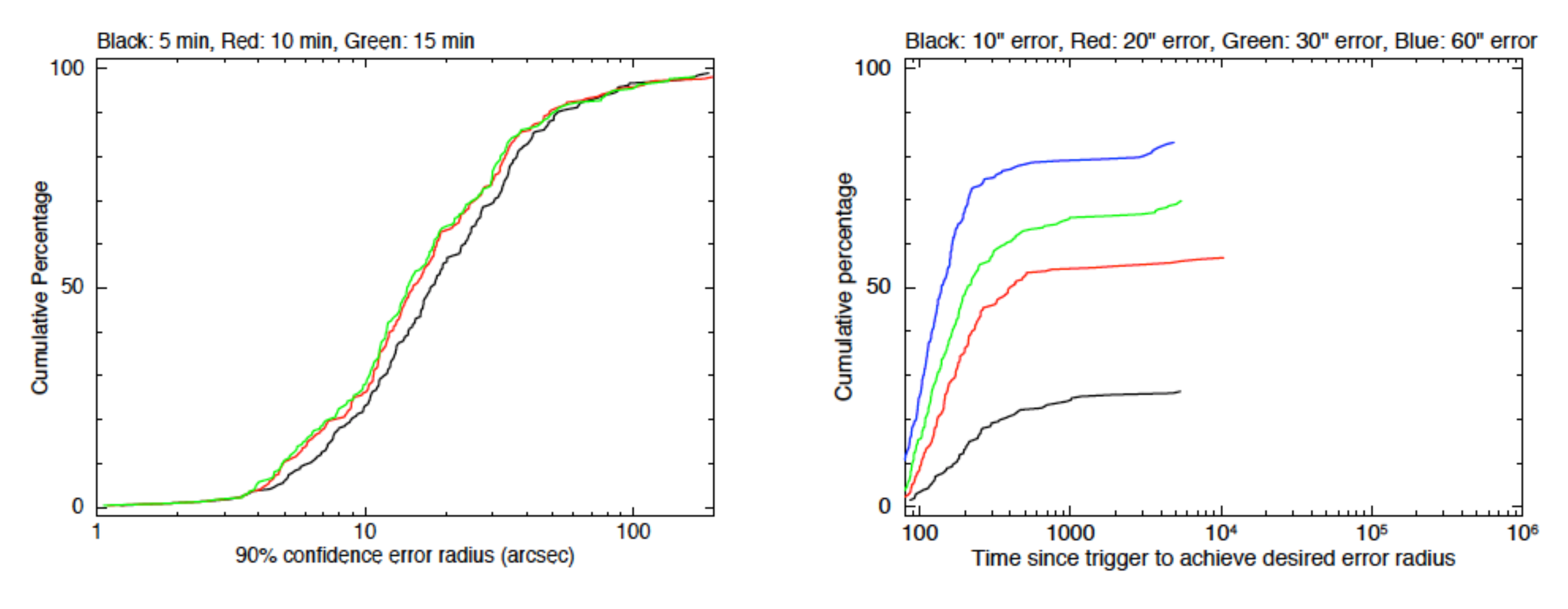}
   \end{center}
   \caption{Left: Cumulative distributions for all GRB afterglows detected of the 90\% confidence MXT position error radii for three different times after the trigger (black: 5 minutes, red: 10 minutes, and green: 15 minutes). Right: The cumulative distributions of various 90\% confidence error radii as a function of time since trigger for all bursts to which SVOM has slewed.}
	\label{fig:locvstime}
	\end{figure}

As can be seen from Fig. \ref{fig:locvstime}, at least 20\% of the GRBs observed by MXT are expected to be localized to a statistical uncertainty of less than 10 arc sec. In order not to degrade the intrinsic localization performance, the systematic residual error between the MXT reference frame and the J2000 coordinate system (achieved through the use of the VT as a “star tracker”, and including the MXT geometrical error, and the J2000 conversion error) shall be smaller than 5 arc sec. This constrains among others the knowledge of the relative bore sight between MXT, the star trackers, and the VT. On the other hand one can see that for 90\% of the GRBs, 15 minutes after trigger, i.e. roughly 10-12 minutes after the MXT started to observe the afterglow, the intrinsic error radius is of the order of 1 arc min, allowing some margin with respect to the [SV-MRR-R7] requirement, provided the instrument misalignments at instrument level, its stiffness, and that the conversion to the J2000 reference frame are not affecting this value too much.

\subsubsection{Sensitivity}
\label{sec:sens}

The sensitivity of the MXT is such that it satisfies the localization requirements at mission level. In fact the localization capabilities of MXT are directly linked to the counts statistics in the early afterglow observations, while the background starts to play a role at later times, when the afterglow is faint. 
We performed some preliminary MonteCarlo simulations in Geant4 to investigate the pnCCD   particle background in the SVOM orbit. We assumed as a first case study a surrounding shield of Cu to refer to the same design adopted for eROSITA, obtaining a non X-ray background level of 9.4$\times$10$^{-3}$ cts cm$^{-2}$ s$^{-1}$ between 0.2 and 10 keV (no SAA passages have been considered for the moment). Work is in progress to design in Geant4 a more realistic configuration based on the use of Al (instead of Cu) plus some internal metal and/or graded-Z coating, and to include the SAA radiation environment among the inputs of Geant4 simulations. On the other hand, the CXB contribution is expected to be of the order of 0.18 cts cm$^{-2}$ s$^{-1}$. 
Under these assumptions on-axis MXT sensitivity is expected to be 10$^{-10}$ erg cm$^{-2}$ s$^{-1}$ ($\sim$5 mCrab, 5 sigma detection level, in 10 s, absorbed flux in the range 0.3-6 keV, for a source with a power law spectrum with photon index=2 and absorption 2$\times$10$^{21}$ cm$^{-2}$). The latter is applicable for short ($<$1000 s) observations, and conforms to [SV-MRR-R5], i.e. it corresponds to about 10$^{-11}$ erg cm$^{-2}$ s$^{-1}$ in 5 minutes. On the other hand, for longer observations, when the background starts to play a role, the sensitivity is expected to be $\sim$2$\times$10$^{-12}$ erg cm$^{-2}$ s$^{-1}$  (5 sigma detection level in 10 ks). We note that those values refer to an on-axis source, and that off-axis vignetting and stray-light will affect these numbers reducing somewhat the sensitivity.

\begin{figure}[ht!]
   \begin{center}
   \includegraphics[width=18cm]{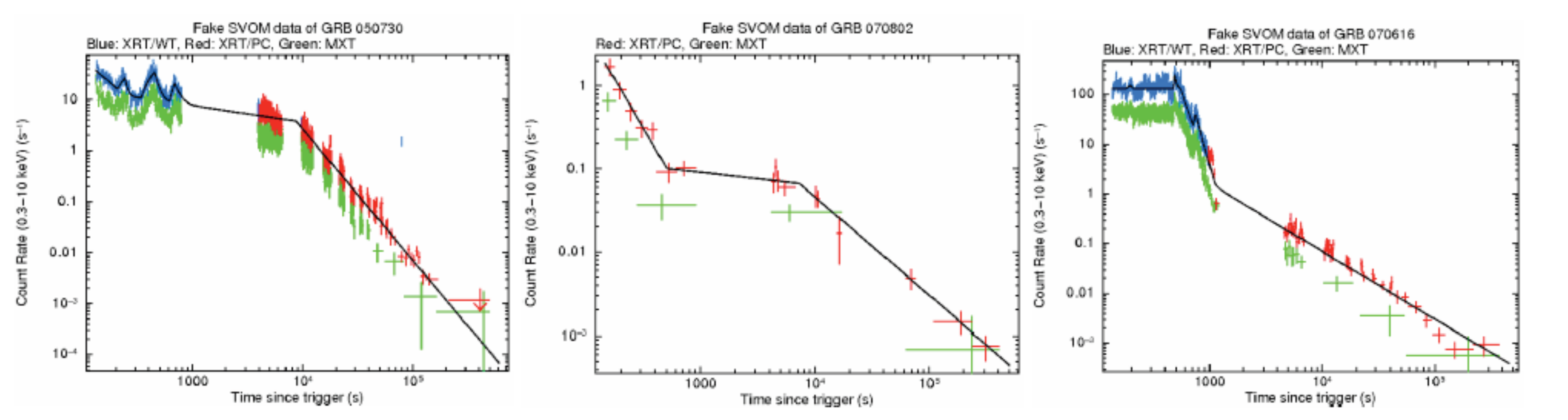}

   \end{center}
   \caption{Simulated MXT GRB X-ray afterglow light curves are shown in green, with the Swift XRT light curves from which they were derived shown in blue and red. The same binning algorithm is used for both instrument's light curves. The first-orbit brightness rank of the afterglows in the Swift sample are: Left 65\%, Centre 21\%, Right 91\%.}
	\label{fig:lc}
	\end{figure}

From the simulations reported in Fig. \ref{fig:lc}, where observations of an average, faint, and a bright X-ray afterglow are shown, one can see that the sensitivity of MXT is such that 

\begin{itemize}

\item all the GRBs detected by ECLAIRs will be detected and localized by MXT, given that they are pointed sufficiently early (1st orbit)
\item even for faint GRB afterglows a detection is expected up to 10$^{5}$ s after the T$_{0}$ (i.e. up to $\sim$18 orbits, provided integration over a few orbits), while after that time the chances of detecting the afterglow are small.

\end{itemize}

\begin{figure}[ht!]
   \begin{center}
   \includegraphics[width=8cm]{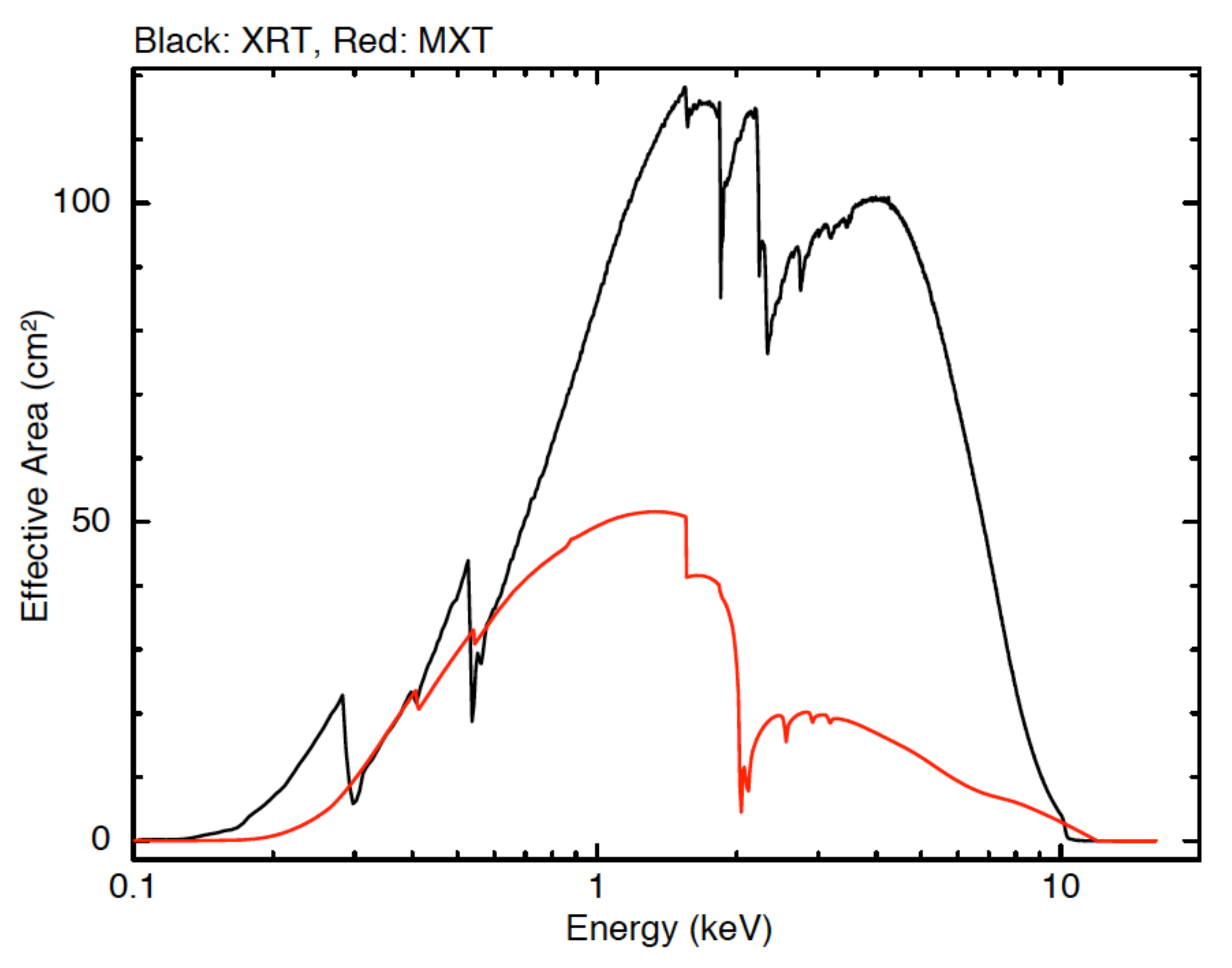}
   \hspace{0.5cm}   
   \includegraphics[width=8cm]{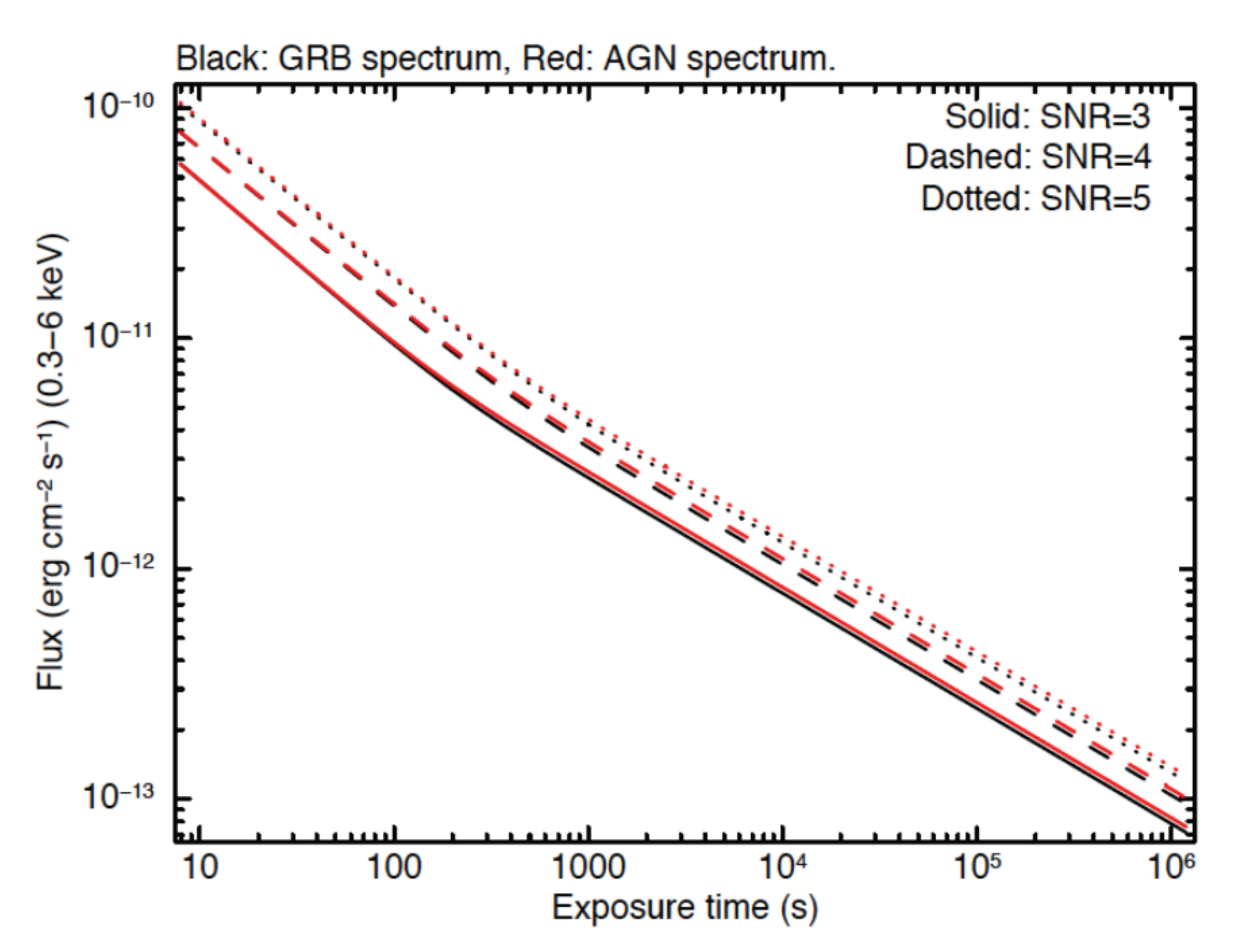}
   \end{center}
   \caption{Left: MXT on-axis effective area (red) shown with the XRT effective area (black). The MXT area curve derived from the \textit{DUO} CCD quantum efficiency, and an assumed 200 nm Al filter. Right: estimated MXT sensitivity as a function of exposure time. The curves show the 0.3-6 keV absorbed flux detection limits for 3, 4 and 5 sigma significance above background as a function of integration time for a detect and background cell size fixed at 80\% of the enclosed PSF. The \textit{DUO} CCD is used with a total 200 nm Al assumed in the beam. The GRB curves assume 82\% vignetting due to typical ECLAIRs position errors, and the typical observed GRB spectrum of Photon Index = 2.0 and N$_{H}$ = 2$\times$10$^{21}$ atoms cm$^{-2}$. The AGN curves assume no vignetting and a typical AGN spectrum of Photon Index = 1.7 and N$_{H}$ = 3$\times$10$^{20}$ atoms cm$^{-2}$.}
	\label{fig:sens}
	\end{figure}

The sensitivity of the MXT as a function of exposure time is shown in Fig. \ref{fig:sens}.

%
%
%

 
\subsubsection{Optical Filter}

The background values derived above do not take into account the possibility of having a bright optical source in the field of view of the MXT. Since X-ray optimised CCDs are also sensitive to IR, optical, and UV light, this will generate an increased background on the detector.
The \textit{DUO} pnCCDs are coated with 100 nm of Al. In addition, for thermal reasons, the MIXS-T MCPs will also be coated in front with an Al film of 60 nm. This will reduce the optical load on the camera during GRB afterglow observation (10$^{-8}$ transmission @ 10 eV), keeping the background at an acceptable level, while having an acceptable transmission at X-ray energies (0.7 @ 500 eV). On the other hand, during non-GRB science observations, bright field stars could be present in the field of view of the MXT, and/or the primary target for VT may be intrinsically optically bright. 
For Swift/XRT a single fixed filter solution has been adopted consisting of 170 nm of polyimide coated on one side with 82.5 nm of Al. Bright optical sources can leak through the XRT optical blocking filter, causing "optical loading" that can complicate or compromise the analysis of the X-ray data.  This implies that objects with a magnitude m$_{V}$ $<$ 8 mag cannot be observed by XRT in photon counting mode, since the data may be poorly calibrated. This shows that bright optical sources could constrain the SVOM pointings, or lower the observation efficiency for MXT.
A different approach has been adopted for the EPIC camera\cite{epic}: a filter wheel with three optical filters has been implemented: thin, medium, and thick. With the thick filter objects in the 0-3 m$_{V}$ can be observed with EPIC/pn, with the medium one ($\sim$10$^{3}$ times less efficient than the thick one), sources up to m$_{V}$ $\sim$8-10 can be observed, while with the thin filter sources fainter than m$_{V} \sim$14 are allowed in the field of view of EPIC.
The XRT filter corresponds roughly to the medium EPIC filter. GRB afterglows rarely exceed magnitude 14, so a solution between the thin and the medium EPIC filters seems appropriate for the MXT fixed filter, as it is the case for the current design. On the other hand, the necessity of a supplementary filter, to reach an attenuation factor e.g. intermediate between the EPIC medium (or XRT) and thick filter shall be evaluated also taking the sensitivity of the VT and its saturation level into account.
Finally, we note that due to the different optical characteristics of EPIC, and XRT, with respect to MXT, their results cannot be scaled directly, and further definition work is needed.  

\subsection{Energy resolution}

\begin{figure}[ht!]
   \begin{center}
   \includegraphics[width=15cm]{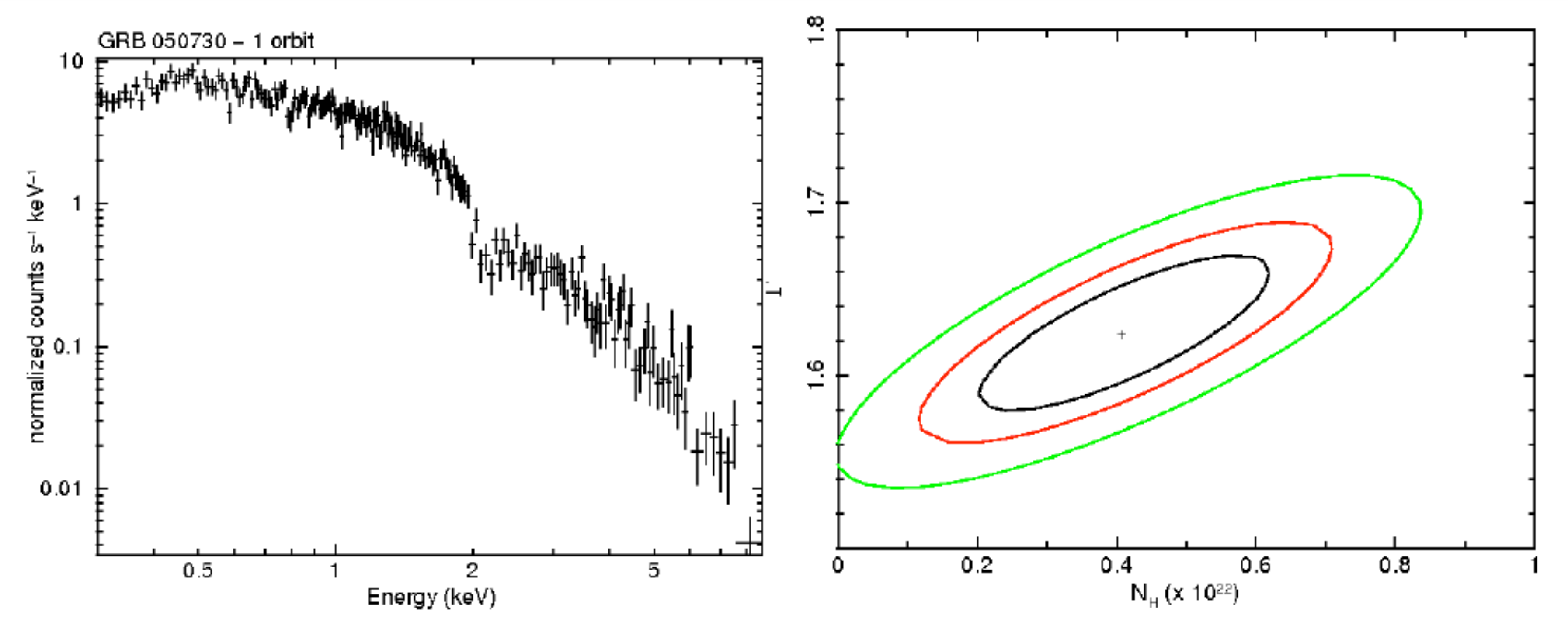}
   \end{center}
   \caption{Left: First orbit background-subtracted simulated MXT spectrum of the Swift XRT burst (GRB 050730) brighter than 65\% of all bursts assuming a 660 second exposure (shorter than the median 1070 seconds exposure in first orbits). Right: 1, 2 and 3 sigma confidence contours of photon index and host galaxy absorption derived from a spectral fit to the spectrum shown on the left, N$_{H}$(ISM) was fixed at the Galactic value and the redshift was fixed at the known z=3.97.}
	\label{fig:sp}
	\end{figure}

As shown in Fig. \ref{fig:sp}, with the standard energy resolution of the adopted pnCCD without any radiation damage ($\Delta$E = 75  eV (FWHM) at 1 keV) MXT can properly characterize GRB afterglow spectra (e.g. the error on the power law slope will be typically of $\pm$4\% (95\% c.l.)). But given the complicated thermal environment and the expected severe radiation environment at 30$^{\circ}$ inclination for the SVOM orbit, at the end of the nominal mission life time (3 years), the energy resolution could be degraded by up to a factor 30\% at Si K (1.8 keV). We note that this estimation is based on the Swift/XRT performance, assuming a cooled (at about -65$^{\circ}$ C) and shielded CCD, and cannot be directly applied to the \textit{DUO} CCD. We stress that in order to reduce the dark current increase and to limit the Charge Transfer Inefficiency (CTI) due to irradiation in space, the cooling and shielding of the CCD are vital functions to be implemented in the MXT design, as well as regular calibration\cite{pagani11}.

\subsection{Pile-up, time resolution and accuracy}
The time resolution of a frame transfer CCD is set by the frame repeat time. However, multiple photons landing in a individual pixel within a frame accumulation time will be misinterpreted as a single photon with the same total energy. This information-destroying effect is known as pile-up, affecting the spatial, spectral, timing and grade distribution of the detected events.
This effect causes distortions in the instrument PSF and difficulties in the energy reconstruction of the events. In order to avoid this undesirable effect, the CCD has to be read out sufficiently quickly. Simulations have shown that given the width of the MXT PSF,
the size of the CCD pixels and the MXT effective area, the 100 ms frame time is adequate to  limit the photon pile-up, since only bursts brighter than about 470 cts/s should be affected by pile-up (at a 10\% level), i.e. maximum 1 per year.
On the other hand the accuracy with which the CCD frame times are known with respect to the satellite TAI system (International Atomic Time) is needed to do precise timing measurements of astrophysical sources. 10 ms is adequate for GRB and non-GRB science (e.g. pulsars, magnetars, etc.). 

\section{Conclusions}
We have shown that the phase A design of the Micrcochannel X-Ray Telescope is well adapted to the scientific requirements of the SVOM mission. Despite the fact the SVOM mission has been implemented on a \textit{mini class} satellite, the coupling of innovative light X-ray optics, based on squared microchannel plates, with a low-noise X-ray camera, allows to build an instrument with an expected sensitivity which is not far from the one of traditional X-ray telescopes with much higher needs in terms of resources.

\section{Acknowledgments}
The authors are deeply indebted to the late Prof. George Fraser for his constant support and invaluable expertise during all the phases of the MXT project. 
JO, AB and PE acknowledge support from the UK Space Agency.

\bibliography{mxt_spie_rev}   
\bibliographystyle{spiebib}   

\end{document}